\documentstyle[aps,epsfig,twocolumn]{revtex}
\begin{document}

\draft
\title{Transport theory with self-consistent confinement
related to the lattice data}

\author
{P.\
Bo\.{z}ek\cite{ja},
 Y.B. He
and J. H\"ufner  }

\address{Institute of Theoretical Physics, Heidelberg University,
Philosophenweg 19, D-69120 Heidelberg, Germany}

\date{\today}

\maketitle

\begin{abstract}
{\rm
 The space-time development of a quark-gluon 
plasma is calculated from a Vlasov equation for the 
distribution function  of quasiparticles with 
medium dependent masses.
 At each space-time point the masses are calculated 
selfconsistently from a gap equation, whose form is 
determined by the requirement that in thermal equilibrium 
and for a range of  temperatures the energy density of the
 quasi-particle system is identical to the one from lattice 
calculations . The numerical solutions of the Vlasov equation 
display confinement. Relations to effective theories like 
that by Friedberg Lee and Nambu Jona-Lasinio  are 
established. \smallskip\\}
\end{abstract}
PACS numbers: 25.75.-q, 12.38.Mh, 12.39.Ba



\section{Introduction}
In view of the possible creation of a quark-gluon-plasma 
(QGP) in high-energy heavy ion reactions at the SPS, RHIC and 
LHC accelerators, descriptions of the space-time evolution 
of the deconfined phase and its transition into the hadronic 
phase are very much called for. Ideally, the transport 
equations should be directly derived from the QCD-Lagrangian, 
a goal which is not yet achieved. Instead a number 
of approximate formulations have been proposed. They are 
either derived from effective Lagrangians (like those by 
Friedberg Lee \cite{book_fl} or Nambu Jona-Lasinio \cite{NJL} (NJL)) 
or they are of Monte-Carlo cascade type \cite{cascade} 
with experimental input like 
string-fragmentation functions and/or cross sections. Both 
approaches have their merits. While the ones based on 
effective Lagrangians permit to study certain aspects of QCD 
(e.g. chiral symmetry) and their manifestation in the space-time
 evolution of a highly excited strongly interacting 
system, the phenomenological approaches are very helpful in 
the interpretation of the data. 

   The approach presented in this paper is closer in spirit to 
the transport theories based on effective Lagrangians.
We propose a Vlasov equation for the evolution of partons, where 
the medium dependent mass is directly obtained from QCD, more precisely from 
the results of lattice calculations.
We concentrate on the transport properties of 
the QGP close to the phase transition, but do not describe 
hadron dynamics. The lattice calculations for QCD at finite 
temperature show a phase transition at a critical temperature 
$T_c$ which separates the regime of the QGP from the one of 
the hadrons. Two aspects are related to the phase transition: 
restoration of chiral symmetry and confinement. They 
manifest themselves in rapid temperature variations around 
$T_c$ of the chiral condensate and the Poliakov loop, respectively.  
Both aspect are contained in the results (often called "data") 
of lattice calculations. 

The basic assumption of our approach is the quasi-particle 
picture, with an effective mass $m$, which in thermal
equilibrium depends on the 
temperature  $T$, and on
space time ($x$, $t$), in the non-equilibrium situation. We treat 
the equilibrium case in Sec. II of our paper and show, how 
$m(T)$ can be obtained from results of the lattice calculations. 
Sec. III  deals with the non-equilibrium case, where 
$m(x,t)$ is calculated from the same gap-equation as $m(T)$. We 
report results of a numerical calculation for the expansion of 
a parton plasma, which shows confinement. The relation to other models 
is given in Sec. IV.

\section{Thermodynamics of the deconfinement phase transition 
in a quasi-particle approach}

\label{sec_2}

The quasi-particle model is one of the most simple approximations to an
interacting many body system. A  system which
consists of particles with definite mass and interactions among them
is replaced by
a system of non-interacting quasi-particles whose masses $m(T)$ are
chosen so that the thermodynamics of the original system is best
approximated. For the case of QCD we make an ansatz for the pressure
density (thermodynamic potential) for the partons \cite{goren}:
\begin{eqnarray}
\label{press_equ}
P_{qp}(m_1,m_2,\dots,T)& = &
\sum_i  g_i \int \frac{d^3p}{(2 \pi)^3} \frac{p^2}{3 E_i(p)}
f_i(E_i(p)) \nonumber \\
& &  - V(m_1,m_2,\dots) \  ,
\end{eqnarray}
where the sum runs over  the parton species $i$ with degeneracy factor 
$g_i$ and where $f_i(E_i(p))$ are Bose or Fermi distribution functions,
which depend on the quasi-particle energies
$E_i(p)=\sqrt{p^2+m_i^2}$.
 $V(m_1,m_2,\dots)$ describes the mean-field contribution to the
dispersion relation \cite{goren}. The potential density $V$
contributes nontrivially  to
the thermodynamic relations   whenever the
masses are  temperature dependent.

The masses $m_i$ appearing in the pressure are phenomenological parameters
which have to be chosen so that the thermodynamic potential
 (\ref{press_equ}) has a minimum,
\begin{equation}
\bigg(\frac{\partial P}{\partial m_i} \bigg)_T=0 \ , \ \ i=1,2,\dots \ \ ,
\end{equation}
leading to :
\begin{equation}
\label{gap_equ}
\frac{\partial V}{\partial m_i} + 
 g_i \int \frac{d^3p}{(2 \pi)^3} \frac{m_i}{E_i}
f_i(E_i) = 0 \ .
\end{equation}
These equations have the form of a gap equation for the masses $m_i(T)$
provided the potential density $V(m_1,m_2,\dots)$ is given.
 Eq. (\ref{gap_equ}) 
 is equivalent to the consistency relations 
\cite{goren,levai}  allowing to obtain 
the energy density from the pressure (\ref{press_equ}), 
in the form as for the ideal gas~:
\begin{eqnarray}
\label{ener_equ}
\epsilon_{qp}(T)& =& \sum_i  g_i \int \frac{d^3p}{(2 \pi)^3} E_i
f_i(E_i) + V(T) \nonumber \\
&= & \epsilon_{kin}(m,T)+V(T) \  ,
\end{eqnarray}
where $V(T)=V(m_1(T),m_2(T),\dots)$.

\begin{figure}[] 
\begin{center}
\includegraphics[width=9cm,angle=0]{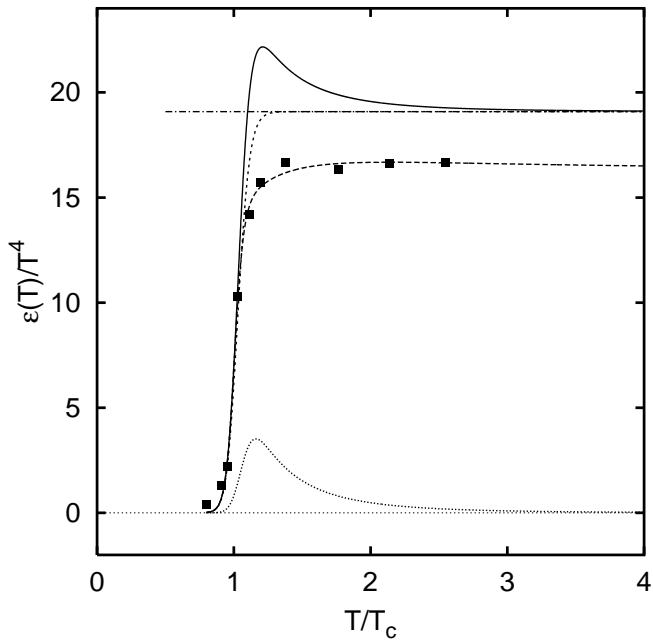}
\caption[]{The energy density of the four-flavor QCD
divided by $T^4$ as a function of the scaled temperature $T/T_c$.
 The points are from the 
lattice data \cite{lattice}. The solid line is the result of the 
quasiparticle model described in the text with the chiral symmetry
restoration (case (b) and solid line in Fig. 2).
 The short-dashed and the dotted 
lines are the corresponding  kinetic
energy density and potential energy density respectively.
The dashed line represents the energy density obtained as a
parameterization
of the data points and leading to the parton mass increasing at high
temperature, (case (a) and dashed line in Fig. 2).
 The dashed-dotted line indicates the Stefan-Boltzmann limit.}
\label{lat_fig}
\end{center} 
\end{figure}

We want to model the deconfinement phase transition of QCD in a
quasi-particle approach and therefore have  to set a criterion how to
determine the quasi-particle masses $m_i(T)$ or equivalently the mean-field
energy density $V(T)$. To date, the most detailed information about the
QCD phase transition comes from the lattice calculations which give
various thermodynamic functions, among them the energy density
$\epsilon_{lat}(T)$. We define  our quasi-particle
approach by the requirement
\begin{equation}
\label{fit_equ}
 \epsilon_{qp}(T)=\epsilon_{lat}(T) \ .
\end{equation}
Since Eq. (\ref{fit_equ}) is only one relation, only one function $m(T)$
can be determined. We therefore have to assume that our quasi-particle
system consists of just one kind of massive partons ($i=1$ in Eqs. 
(\ref{press_equ}) to (\ref{ener_equ})). We will assume a Boltzmann type
distribution function $f(E,T)=\exp^{-\beta E}$.
The right hand side in Eq. (\ref{fit_equ})
being given, this equation is a constraint on the unknown
function $m(T)$ in $\epsilon_{qp}(T)$.
We note that  the quasi-particle model constrained by Eq. (\ref{fit_equ})
 describes exactly the energy density of lattice QCD,
but may reproduce other thermodynamic function only approximately 
\cite{goren,levai} which is not a surprise since lattice QCD is not
a quasi-particle gas. 
The functional form for $m(T)$ can be obtained by
solving the differential equation
\begin{eqnarray}
\label{diff_equ}
\frac{dm(T)}{dT}&= &\bigg({\frac{d\epsilon_{lat}(T)}{dT}-\frac{\partial 
\epsilon_{kin}(m,T)}
{\partial T}}\bigg) \nonumber \\
& & /\bigg({\frac{\partial \epsilon_{kin}(m,T)}
{\partial m}+\frac{dV}{dm}}\bigg) \ ,
\end{eqnarray}
which is obtained from Eq. (\ref{fit_equ}) by differentiating both sides
with respect to $T$ and
where ${dV}/{dm}$ is given by the gap equation (\ref{gap_equ}).
Eq. (\ref{diff_equ}) is a first order differential equation which determines 
$m(T)$ for a given energy density $\epsilon_{lat}(T)$ and for an initial
value $m(T_0)$. Then $V(T)$ can be obtained from
$V(T) = \epsilon_{lat}(T)- \epsilon_{kin}(m(T),T)$.
 The lattice data are normalized so that
$\epsilon_{lat}(T)
\rightarrow 0$ for $T \rightarrow 0$. It requires that $V(T)\rightarrow 0$
at low temperatures. 
If $m(T)$ is not obtained from Eq. \ref{diff_equ} but is
 given from some other considerations,
the gap equation can be integrated to obtain the potential $V(T)$~:
\begin{equation}
\label{intV_equ}
V(T)=-g \int_{m(T_0)}^{m(T)}  dm
 \int \frac{d^3p}{(2 \pi)^3} \frac{m}{E(p)}
f(E(p)) +V(T_0)\  ,
\end{equation}
where $g=\sum_i g_i$.

We apply the above methods to the lattice data from \cite{lattice}
shown in Fig. \ref{lat_fig}.  These data
 are calculated for 4 flavors, corrected
for finite lattice size and extrapolated to massless fermions.
 We use  $g=62.8$ corresponding to the noninteracting limit of
 QCD with $4$ massless flavors.
We  draw the attention to the fact that the lattice data in
Fig. \ref{lat_fig}
for  $\epsilon_{lat}(T)/T^4$  do not approach
the Stefan-Boltzmann limit for $T \gg T_c$.
This may have two reasons.
\begin{itemize}
\item [(a)] The parton mass $m(T)$ never approaches the chiral limit
$m=0$. Perturbative arguments (whose validity is questionable around
$T_c$) suggest that $m(T)\sim T$ for large $T$. If we attribute the
discrepancy between $\epsilon_{lat}(T)/T^4$ and the Stefan-Boltzmann
limit to this reason, one finds the minimal mass $m(T)\simeq 2.1 T_c$
right above $T_c$ and $m(T)\simeq 1.1 T$ for large $T$.
\item [(b)] Chiral restoration requires
 that $m(T)\rightarrow 0$  above the phase
transition at least for the fermions. Then  the deviation from the
Stefan-Boltzmann limit has to be attributed 
to a mechanism which is outside the scope of the
quasi-particle approach.
\end{itemize}
\begin{figure}[] 
\begin{center}
\includegraphics[width=9cm,angle=0]{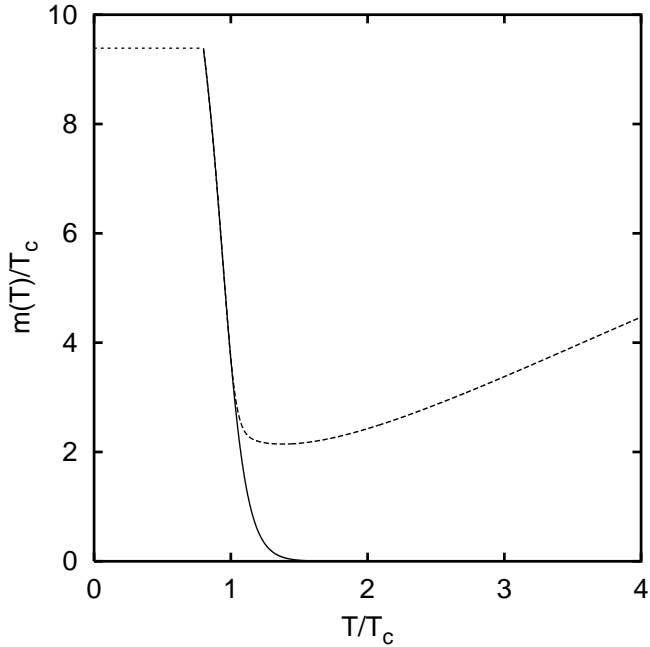}
\caption[]{The dashed line represents the 
temperature dependence of the quasiparticle mass 
as obtained from the lattice data (case (a)).
The short dashed line represents the extrapolation to the low temperature
region. It is taken to be constant, and suffices to effectively confine
the partons (an absolute confinement would require that the mass goes
to infinity at low temperatures). The  solid line represents the 
assumed temperature dependence of the parton mass with zero limit at
high temperature (case (b)).}
\label{mass_fig}
\end{center} 
\end{figure}
We will explore both possibilities and call the corresponding masses
$m_a(T)$ and $m_b(T)$ respectively.
First we solve the Eq. (\ref{diff_equ}) to obtain $m_a(T)$ for the case
(a),
where we set $m_a(T_0)=9.5T_c$ for $T_0=0.8T_c$.
For the case (b) we take $m_a(T)$ obtained in (a) for $T < T_c$ and
extrapolate it at $T>T_c$ so that $m(T) \rightarrow 0$ at large $T$
(Fig. \ref{mass_fig}).
From this new functional form $m_b(T)$ we obtain $V_b(T)$ and
$\epsilon_b(T)$ via Eq. (\ref{intV_equ}).
Fig. \ref{mass_fig} shows the obtained temperature dependence of the parton mass for
the two cases. While the two functions $m(T)$ coincide
in the region of the confinement transition (below $T_c$) they differ
dramatically for 
$T>T_c$. The mass $m_a(T)$ increases linearly with $T$
for large temperatures while $m_b(T)$ is set to reach the chiral limit
at high temperatures.
 Fig. \ref{lat_fig}  also shows also the energy density
corresponding to the two cases. We observe that in the case (a) 
 $\epsilon_{qp}(T)$ follows  the lattice points as it should.
 In particular it has the same high temperature
limit which is different from the Stefan-Boltzmann limit.
\begin{figure}[t] 
\begin{center}
\includegraphics[width=9cm,angle=0]{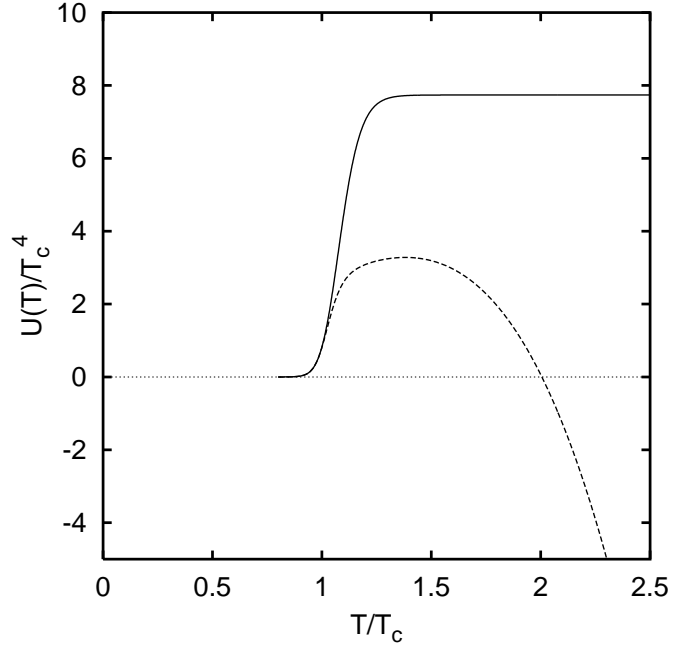}
\caption[]{The potential energy density $V(T)$ as a function of the
temperature for the case (a) and (b) of the temperature dependence of
the mass, dashed and solid line respectively.}
\label{pot_fig}
\end{center} 
\end{figure}
\noindent The energy density for the case (b) is different for $T>T_c$ and
approaches the Stefan-Boltzmann limit. 
 Fig. \ref{lat_fig}  also shows the densities for the  kinetic energy
 $\epsilon_{kin}(m(T),T)$ and for the potential energy $V(T)$
 for the case (b).
Fig. \ref{pot_fig}
 shows the potential density functions $V(T)$ for both  cases. 
In the case (b), one can identify the large temperature
limit of $V_b(T)$ with the bag constant $B$, since $V_b(T=0)=0$.
 Indeed for large $T$ (in the
deconfined phase) Eqs. (\ref{press_equ}) and (\ref{ener_equ}) take the
form of the free massless gas with a bag constant
$B=V_b(T=\infty)$. 
For the case (a) one can {\it define} the bag constant as the
value of the potential density $V_a(T)$ for the temperature where the mass
 $m_a(T)$ is the smallest (slightly above $T_c$). 
We find
\begin{equation}
B^{1/4} =\left\{ \begin{array}{rcll} 
1.35T_c & = & 240 \ \textrm{MeV} & \ \ \textrm{ for case (a)} \\
1.67T_c & = & 300 \ \textrm{MeV} & \ \ \textrm{ for case (b)} 
\end{array} \right.
\end{equation}
for a value of $T_c=180$ MeV corresponding to four-flavor QCD.
 The order of magnitude of the bag constant is correct, but in
order to compare it to the usual bag constant ($B^{1/4}=135-200$ MeV)
 one should use the lattice
results for the energy density and
the critical temperature of the two-flavor QCD as an input for the 
quasi-particle parton model.

In order to distinguish between the two solutions for $m(T)$ we
investigate the temperature dependence of the chiral condensate
in the quasi-particle picture
\begin{equation}
\label{chi_equ}
\langle {\bar \Psi} \Psi  \rangle_{qp} (T) = 
\langle {\bar \Psi} \Psi  \rangle_{vac} (T) + g_{q{\bar q}}
\int \frac{d^3p}{(2 \pi)^3}\frac{m}{E} f(E) \ ,
\end{equation} 
where $g_{q{\bar q}}$ counts  fermion and anti-fermion
 degrees of freedom 
and the vacuum part of the chiral condensate is given by an expression
including a cutoff in momentum
\begin{equation}
\langle {\bar \Psi} \Psi  \rangle_{vac} (T) =  g_{q{\bar q}}
\int_{|p|<\Lambda} \frac{d^3p}{2(2 \pi)^3}\frac{m}{E}  \ .
\end{equation} 
\begin{figure}[b] 
\begin{center}
\includegraphics[width=9cm,angle=0]{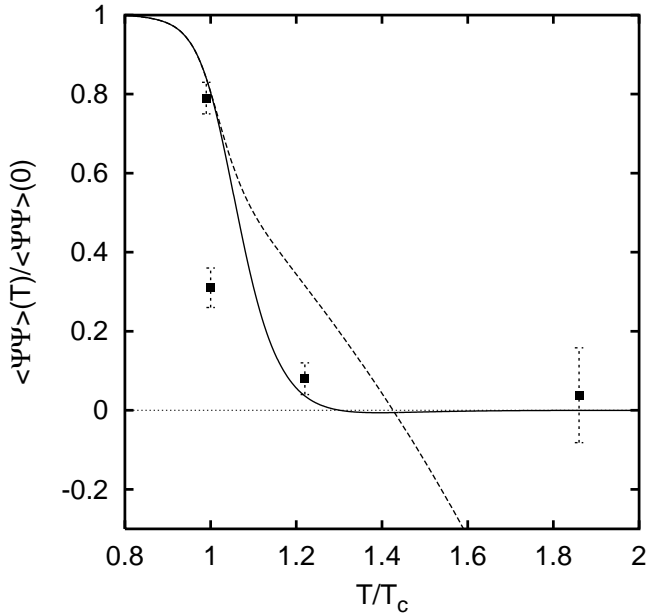}
\caption[]{The  chiral condensate as a function of the temperature
 for the case (a) and (b) of the temperature dependence of the mass,
 dashed and solid line respectively. The data points are for
the lattice data in the four-flavor QCD \cite{kar}.}
\label{ch_fig}
\end{center} 
\end{figure}
\noindent
The value of the cutoff $\Lambda=430$ MeV is fixed to reproduce the zero
temperature value of the chiral condensate which we take 
twice the usual value
\begin{equation}
\langle {\bar \Psi} \Psi  \rangle_{vac} (0) =  2 (250 \textrm{MeV})^4 \ ,
\end{equation} 
because we are modeling the  four-flavor lattice QCD.
Fig. \ref{ch_fig} shows a comparison between the condensate functions for the
quasi-particle picture and the lattice data. This comparison clearly
favors case (b), where $m(T)\rightarrow 0$ above $T_c$. In
what follows, we will mainly work with this solution.


Before we treat the nonequilibrium case
we apply the gap  equation to the case of finite density  and 
calculate 
the density dependence of the parton mass.
We  introduce variables which will be also useful for the
discussion of the nonequilibrium case. 
First let us write the gap equation in the form
\begin{equation}
\frac{dV}{dm}=V'(\rho)\frac{d\rho}{dm}=-\rho \ ,
\end{equation}
where we define the scalar density
\begin{equation}
\rho=
 g \int \frac{d^3p}{(2 \pi)^3} \frac{m}{E(p)}
f(E(p)) \ .
\end{equation}
\begin{figure}[b] 
\begin{center}
\includegraphics[width=9cm,angle=0]{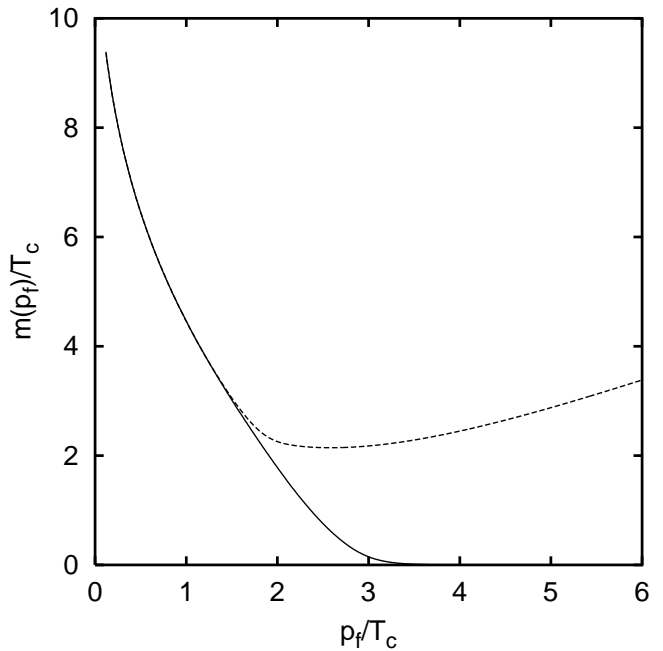}
\caption[]{The  quasiparticle mass  as a function of the scaled 
Fermi momentum $p_f/T_c$. The solid and  dashed lines correspond to two
different temperature dependences of the parton mass in Fig. 2.
}
\label{fd_fig}
\end{center} 
\end{figure}
\noindent
It allows us to use the functions $V(\rho)$ and $m(\rho)$ instead of $V(T)$
and $m(T)$ for the parameterization of the potential energy density and
the mass respectively.
Assuming that the energy density depends on temperature $T$ and chemical
potential $\mu$ only through  $\rho(T,\mu)$ we can generalize
the finite temperature case also to finite density. In principle the
potential $V(\rho)$ could depend on  other quantities,  e.g. the
baryon density. Any such more general case cannot be discussed using
only the lattice data at finite temperature.
As a support for our assumption we note that for the
NJL model the potential density $V$  depends only 
on the density $\rho$ (Sec. \ref{other_sec}).

We  write the gap equation (\ref{gap_equ}) at finite density
\begin{equation}
\label{gapden_equ}
\frac{dV}{dm}=- g_f \int \frac{d^3p}{(2 \pi)^3} \frac{m}{E(p)}
\Theta(p_f-|p|) \ ,
\end{equation}
where $g_f$ counts the number of fermion degrees of freedom.
Note that here we are using the Fermi distribution for the fermions
at finite density and zero temperature and $V(m)$ from the lattice data.
Fig. \ref{fd_fig}
 shows the mass of the partons as a function of the Fermi momentum $p_f$.
The behavior is similar as in the finite temperature case.
For low density the mass increases leading to the effective confinement.
At high density the mass is proportional to $p_f$ for the case (a) and
goes to zero in the case (b).

\section{Transport theory of the effective confining model}

In this section we will discuss the nonequilibrium evolution of the 
parton densities. In the semiclassical limit the collisonless 
 plasma of quasi-particle partons
 can be described by the  Vlasov equation for the phase-space
distribution function $f(x,t,p)$:
\begin{eqnarray}
\label{vlasov_equ}
\partial_t f(x,t,p)+\frac{ p}{E(p,x,t)} \nabla_x f(x,t,p)
& & \nonumber \\
- \frac{m(x,t)}{E(p,x,t)} \nabla_x m(x,t) \nabla_p f(x,t,p)=0 \ .
\end{eqnarray}
Eq. (\ref{vlasov_equ}) has to be supplemented by an equation for the
space-time dependent mass  $m(x,t)$ .
A sufficient condition for the requirement, that the Vlasov equation
describes the same physics at thermal equilibrium as presented
in the
previous section is that $m(x,t)$ satisfies the same gap equation  
\begin{equation}
\label{gapgen_equ}
 \frac{dV}{dm} = -  g  \int \frac{d^3p}{(2 \pi)^3}
\frac{m(x,t)}{E(p,x,t)} f(x,t,p) = -\rho(x,t) \ ,
\end{equation}
where the thermal distribution function $f(E)$  in Eq. (\ref{gap_equ})
has been replaced by the nonequilibrium solution of the Vlasov
equation $f(x,t,p)$ and where $V(m)$ is the same as in equilibrium.
 The solution of the nonequilibrium gap equation is 
then a function $m(x,t)$ of space and time.
Eq. (\ref{gapgen_equ}) is however not the most general equation which reduces
to the equilibrium finite temperature gap Eq. (\ref{gap_equ}),
but one can add to it  terms which depend on the
spatial and time derivatives of $m(x,t)$.
We will come back to this question in Sec. \ref{other_sec}.
Here we note that the choice in Eq. (\ref{gapgen_equ})
 guarantees that the total energy of the system
\begin{eqnarray}
E(t)& =& g \int d^3x \int \frac{d^3p}{(2 \pi)^3} E(p,x,t)
f(x,t,p) \nonumber \\
 & & +\int d^3x / V(\rho(x,t)) 
\end{eqnarray}
is  conserved by the evolution according to the Vlasov
Eq. (\ref{vlasov_equ}), i.e. the energy of the system is constant.

We have numerically solved  the Vlasov equation together
with the gap equation for the two cases of the functional dependence of
the mass $m(T)$ on the temperature discussed in Sec. \ref{sec_2}, but 
most of the results shown relate to the case where the chiral condensate
vanishes at high temperatures (case (b) in the previous section).
 We have used the test particle method for the solution of the Vlasov
equation, i.e. we have made the ansatz
\begin{equation}
f(x,t,p)=\sum_{j=1}^{N}\delta^3(x-x_j(t))\delta^3(p-p_j(t)) \ ,
\end{equation}
where the trajectories $x_j(t)$ and $p_j(t)$ of the $N$ test particles
 satisfy Hamilton's
equations with
\begin{equation}
H(x,p)=\sqrt{ p^2+m^2(x,t)} \ .
\end{equation}
The initial conditions $x_j(0)$ and $p_j(0)$ are chosen so that
given  initial densities for  matter and momentum  are reproduced.
The initial density is chosen spherically symmetric. Also in the 
solution of the
 gap equation  the spherical symmetry is imposed by angle averaging. 

\begin{figure}[b] 
\begin{center}
\includegraphics[width=9cm,angle=0]{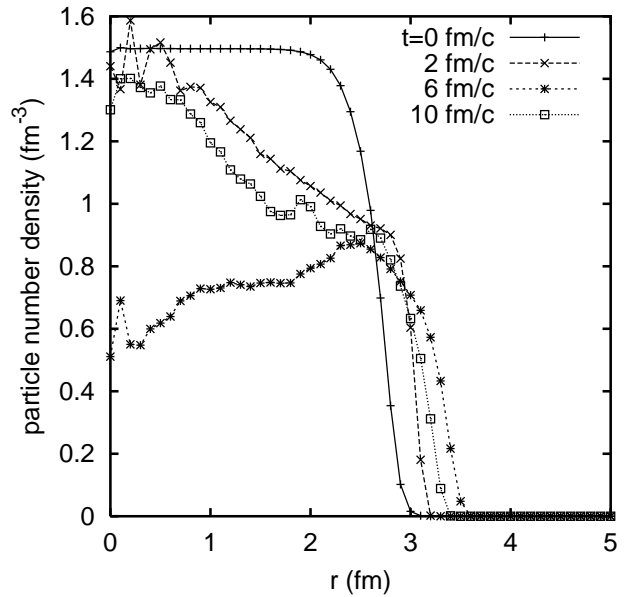}
\caption[]{The  parton density distribution 
 at different times,
as obtained from the nonequilibrium evolution of the 
initial fireball ($t=0$).}
\label{den_fig}
\end{center} 
\end{figure}
\begin{figure}[b] 
\begin{center}
\includegraphics[width=9cm,angle=0]{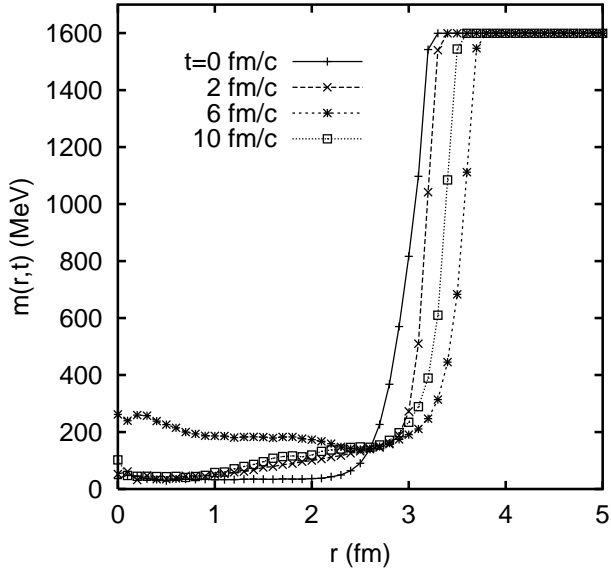}
\caption[]{The  parton mass distribution at different times,
as obtained from the nonequilibrium evolution of the
fireball.}
\label{masstime_fig}
\end{center} 
\end{figure}

At time $t=0$ the system is described by the density profile shown in
Fig. \ref{den_fig} and with a momentum distribution corresponding to a
temperature $T=1.3T_c=180$ MeV. Fig. \ref{den_fig} shows the density for
different times. The system expands for $t=2$ and $6$ fm/c, but then comes
back ($t=10$ fm/c). The dependence of $m(x,t)$ on the radius at different
times is shown in
Fig. \ref{masstime_fig}. As expected
 the mass is small in the interior and
increases towards the surface. This increase is responsible for the
confinement.   Indeed one can show that the equation
 of motion of one particle in the mean-field of the other particles
conserves the  energy of the particle $\sqrt{p^2+m^2}$. Thus
a particle cannot leave the region of deconfined plasma, if
its initial momentum $p$ satisfies:
\begin{equation}
\label{conf_equ}
p^2 < m_{vac}^2-m^2 \ ,
\end{equation}
where $m_{vac}$ is the parton mass in vacuum and $m$ is the initial 
selfconsistent mass of the parton inside the plasma. The vacuum mass
should be 
infinite, if the confinement is absolute. In our calculation we take 
the vacuum mass equal to $\sim 9.4 T_c$
 (see Fig. \ref{mass_fig}), which effectively confines the
 partons at the temperatures discussed here. 
 Thus the partons cannot leave the hot fireball if their initial 
momentum is smaller than $\sim 9.4 T_c$,
 which is the case for most of the partons 
at our initial temperature. 

\begin{figure}[t] 
\begin{center}
\includegraphics[width=9cm,angle=0]{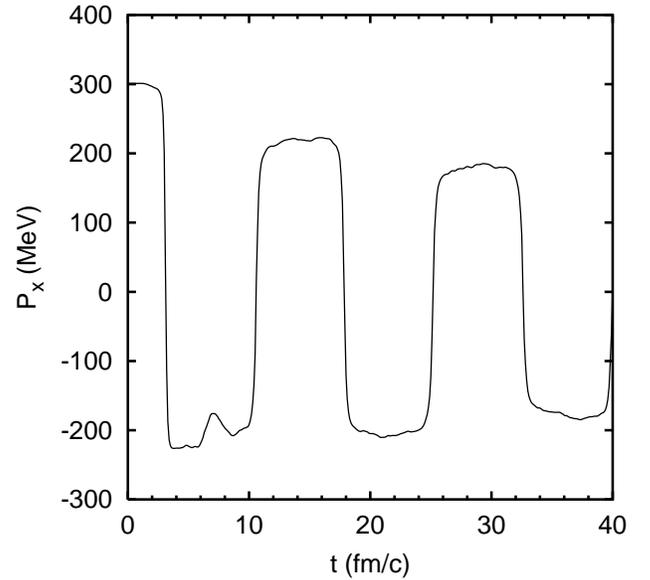}
\caption[]{The component $p_x$ of the momentum of a 
particular test-particle
 taken from the
simulation of the time evolution of the region of deconfined plasma.}
\label{px_fig}
\end{center} 
\end{figure}

\begin{figure}[h] 
\begin{center}
\includegraphics[width=9cm,angle=0]{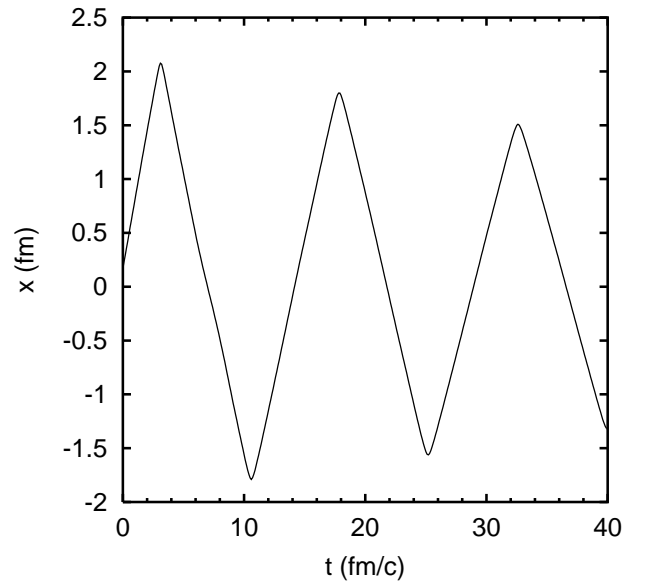}
\caption[]{The  $x$ coordinate of the same test-particle
as in Fig. \ref{px_fig}.}
\label{x_fig}
\end{center} 
\end{figure}

 Figs. \ref{px_fig} and \ref{x_fig}  show 
the time development of the momentum $p_x$ and the coordinate $x$ of
a particular test particle.
 The particle oscillates between 
the borders of the fireball and at the border its momentum is reduced and 
eventually reversed by the action of the force:
\begin{equation}
\frac{dp}{dt}=-\frac{m}{E}\nabla_x m \ .
\end{equation}

In contrast to the simple particle motion, partons traveling in bunches
may leave into the vacuum region, since
 the internally created field leads
to small masses inside the bunch. This possibility is excluded in our
method of 
\noindent
solution, since we require at each time step that the system stays 
  spherically symmetric. This mechanism then
 generates     collective vibrations of the surface of the
 fireball, when particles are trying to leave the hot region 
\begin{figure}[b] 
\begin{center}
\includegraphics[width=9cm,angle=0]{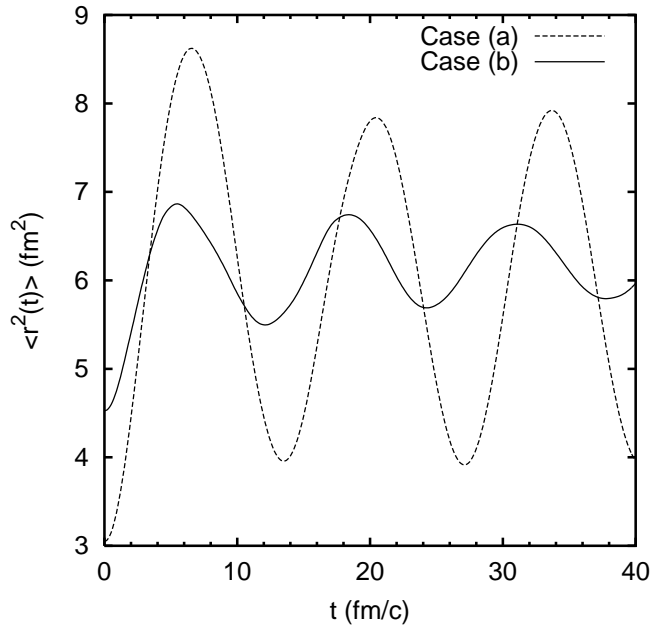}
\caption[]{The time evolution of the mean square radius of the region of
deconfined plasma for the case (a) and (b), dashed and solid  line
respectively}
\label{r2_fig}
\end{center} 
\end{figure}
\noindent simultaneously.
 Fig. \ref{r2_fig}  shows the time dependence for the mean
 square  radius of the fireball~:
\begin{equation}
\langle r^2(t) \rangle =  g\int d^3x \ x^2   \int \frac{d^3p}{(2 \pi)^3} 
f(x,t,p) \ .
\end{equation}
The oscillation which can be seen on the figure reflect the collective 
monopole oscillations of the parton density, with  partons  leaving
the fireball and   reflected back when their mass grows.
Analogous collective oscillations have been observed in the Vlasov
evolution of the nucleon in the Friedberg-Lee model \cite{giessen}.

\begin{figure}[b] 
\begin{center}
\includegraphics[width=9cm,angle=0]{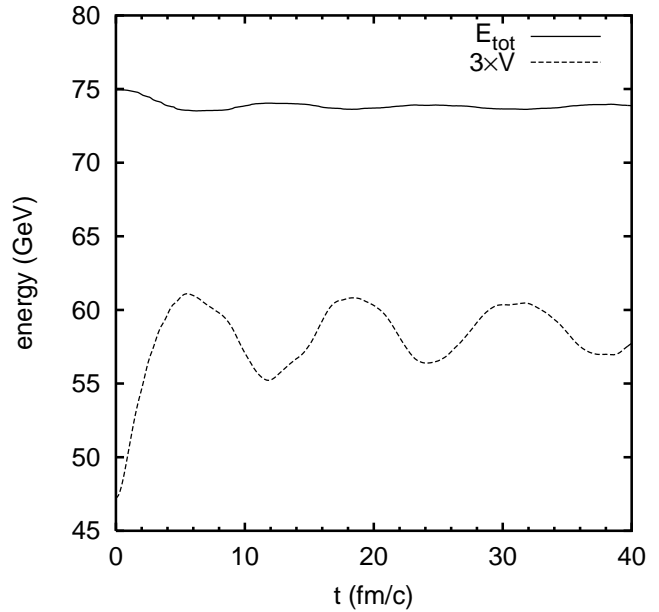}
\caption[]{The time dependence of the total and potential energy 
of the parton plasma, solid and  dashed lines respectively}
\label{ener_fig}
\end{center} 
\end{figure}

As the volume of the
 fireball  oscillates,  its potential energy will also oscillate 
 growing for large volumes and decreasing when the system is compressed.
The period of the collective oscillations is basically determined 
by the time which a particle needs to travel from one border of the
fireball to the other (Fig. \ref{x_fig}). For  massless
deconfined partons (case (b) in Sec. \ref{sec_2}) this time is
twice the radius of the system divided by the speed of light.
The total energy is of course conserved (see Fig. \ref{ener_fig}),
 to the accuracy of 
our numerical solution, and the kinetic (and then also the potential energy)
 has oscillations of the same period  as the oscillations of the 
mean square radius in Fig. \ref{r2_fig}.

We mention the  possibility of  fragmentation
 of the fireball into smaller pieces, each of them having 
large parton density inside and thus a 
small mass of partons. This fragmentation
mechanism cannot be studied in our spherically symmetric mean-field theory.
It would require a description including the fluctuations of the density.

We state the main result of this section: the Vlasov equation together
with the gap equation based on lattice data shows confinement.
For not too high initial temperatures the fireball stays compact. 
Of course  the surface of the fireball
 can oscillate, but the system remains bounded. This
behavior is in contrast to the free expansion of the parton gas, where the
parton fireball would start to expand and cool down faster.
In our picture the partons are confined, since no hadronization is included
(see Fig. \ref{r2_fig}).

\section{Relation to other effective models}
\label{other_sec}

\begin{figure}[b] 
\begin{center}
\includegraphics[width=9cm,angle=0]{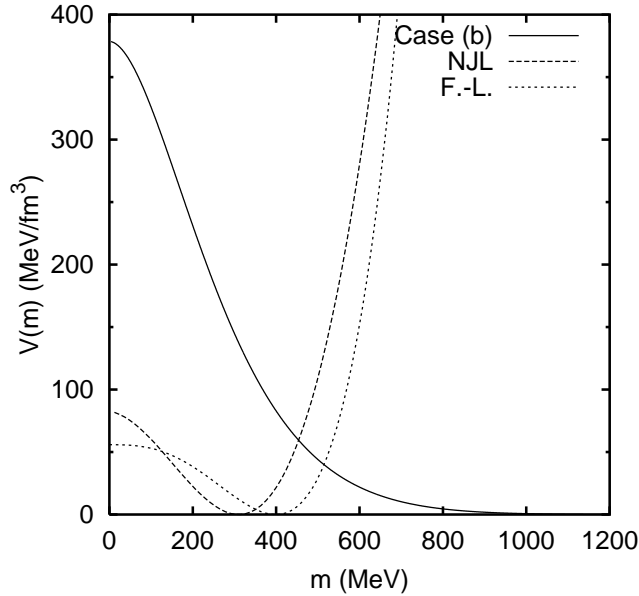}
\caption[]{The potential energy density $V(m)$ as a function of the mass
as extracted from lattice QCD 
(solid line), for the two-flavor NJL model (dashed line)
and for the Friedberg-Lee model (dotted line).
We used $T_c=140$ MeV  in defining $V(m)$.
The large value of the bag constant $B=V(m=0)$ for the potential density 
from the lattice data can be due to a larger number of degrees of freedom
in the four-flavor model. }
\label{pote_fig}
\end{center} 
\end{figure}

The effective models of the QCD using parton degrees of freedom
  are mostly restricted to the fermion
 sector. Thus NJL type models \cite{NJL} have only fermionic degrees of
 freedom with four-fermion interaction. At high temperature and/or
 density the quarks are massless or almost and at low densities due to
 the nonzero quark-condensate they acquire a finite mass with value
 around $350$ MeV. The quarks are not confined in this theory.
 The NJL gap equation
 for the quark mass can also be written in the form of 
Eq. (\ref{gap_equ}) where $dV/dm$ is defined by the expression
\begin{equation}
\frac{dV}{dm} \equiv \frac{(m-m_0)}{2G} - 6 N_f
\int_{|p|<\Lambda} \frac{d^3p}{(2\pi)^3} \frac{m}{E} \ ,
\end{equation}
where $G$ is the four fermion coupling constant, $\Lambda$ is the
infrared cutoff, $m_0$ is the current quark mass and $N_f$ is the number
of flavors. Fig. \ref{pote_fig} shows a comparison between the potential
extracted from lattice calculations and the one from the NJL model.
The difference is twofold: 
(i) Shape: While the potentials for Friedberg-Lee and NJL model show a
clear minimum at the position of the constituent mass in the vacuum, the
potential from our approach drops monotonically to zero, reflecting the
confinement (infinite vacuum mass).
(ii) Magnitude: At $m=0$ the potential from our approach differs by about
a factor $5$ from the other approaches. This may be partly due to the
different numbers of degrees of freedom. While Friedberg-Lee and NJL
refer to a two-flavor quark theory (no gluons), the lattice calculation
is performed for $N_f=4$ and gluons.
The remaining difference may be due to quantitative difference between
the four-flavor and the two-flavor QCD, which goes beyond a simple
rescaling of the number of degrees of freedom.

The Friedberg-Lee model describes fermions coupled to a scalar field
$\sigma(x,t)$ which plays the role of an effective mass and whose
dynamics is driven by a potential $U(\sigma)$.
Of course as in the NJL model the degrees of freedom are restricted to
the fermions. The Lagrangian of the Friedberg-Lee model can be written as:
\begin{equation}
\label{fl_equ}
{\cal L}={\overline \Psi}(i\gamma^\mu\partial_\mu)\Psi -
g \sigma {\overline \Psi}\Psi +\frac{1}{2}\partial_\mu\sigma\partial^\mu
\sigma-U(\sigma) \ ,
\end{equation}
where the fermion field operator $\Psi$ also carries the flavor and color
indices. 
The effective quark mass $g \sigma$ can include also a contribution from
the current quark mass. 

When the expectation value of the $\sigma $ 
field is infinity, the quarks are confined. The same is  also
effectively true
if the vacuum expectation value of the $\sigma$ field is very large.
The gap equation for the quark mass $\sigma$ is given by the classical
equation of motion for the $\sigma$ field:
\begin{equation}
\Box \sigma + U^{'}(\sigma)= - g < {\bar \Psi}\Psi > (x,t) \ .
\end{equation}

In the case of homogeneous systems $\Box \sigma=0$ (e.g. in the
mean-field thermodynamics)
the Friedberg-Lee gap equation is equivalent to the gap equation used in
Sec. \ref{sec_2} if $V(m)=U(\sigma)$.
In particular the thermodynamical energy density discussed in
Sec. \ref{sec_2} can be reproduced in the Friedberg-Lee model 
if not for the neglect of the gluon degrees of freedom. The difference
between our approach and the one by Friedberg and Lee rests in the
choice of the potential. While they assume certain forms,
we let $V(m)$ be determined by the lattice data.
One should note that in the homogeneous systems the confining
Friedberg-Lee model, with the fermion mass $m=\kappa(\sigma)$
being a function of the field $\sigma$, is also equivalent to our
approach after a change of variables $\sigma \rightarrow m$.
The infinite value of the fermion mass  means simply in
the language of Eq. (\ref{fl_equ}), that the 
vacuum gap equation $dV/dm=0$ has a solution at $m=\infty$.

 The kinetic term 
for the $\sigma $  makes a difference for the case of nonequilibrium 
or nonhomogeneous systems. In the dynamical evolution of a
nonequilibrium system in the Friedberg-Lee model, the $\sigma $ field
is another dynamical field not related to the local value of the
scalar density $\rho$ \cite{giessen}. The inclusion of the kinetic term
for the $\sigma $ field however leads to the problem that the value of
the $\sigma$ field can go negative. The potential $U(\sigma)$ cannot be
extracted for the negative values of $\sigma$ from the lattice
data. Moreover in the cases when the mass of the fermion field becomes
negative, its evolution cannot be described by a semiclassical Vlasov
equation.

\section{Discussion}

The description of the deconfinement transition in heavy ion
collisions is  a very important theoretical problem. A realistic
description could allow to define which observables are relevant 
for the observation of the quark-gluon plasma formation. A
dynamical simulation is wished for in order to extract the properties of
the plasma from the experimental data. Any such approach
 meets difficulties in the description of the confinement and of 
the hadronization
transition when the temperature of the deconfined region drops down.
In the present work we have addressed only a part of this program,
namely the influence of the confinement on the dynamics of partons.

The confinement of partons below $T_c$ is described in a quasi-particle
gas model by the increase of the parton mass at low energy densities.
 The temperature dependence
of the parton mass is extracted from the lattice data. 
In the  range
of temperatures analyzed ($0.8T_c <T< T_c$) the mass of the partons
increases to values which effectively confine the quarks and gluons.
  The requirement of the restoration of the chiral symmetry at high 
temperatures fixes the temperature mass dependence for $T>T_c$.

The formalism is generalized to the nonequilibrium case.
The time development of a deconfined
fireball  studied in a Vlasov equation 
is different from what is  usually discussed in the literature,
in that it shows confinement. The expansion of the system
is forbidden, but instead
we observe collective oscillations of the parton plasma.  The time scale
of these oscillations in the collisonless plasma is determined by the
size of the deconfined region. This time scale should be compared to the
hadronization time in order to determine if the oscillation could
develop.

The increase of the parton mass has implication also for the 
hydrodynamical model of the plasma expansion.
The slowing down of the hydrodynamic expansion is observed  in
 simulations of systems with first order phase transition
\cite{sh} or in a hydrodynamical calculation with the NJL model \cite{copen} .
The use of a confining mass in the hydrodynamical model would 
stop the expansion.
Further expansion of the fireball is possible only after
hadronization. 

If the hadronization takes place mainly  
at the surface of the deconfined phase
then the dynamics of the system could be described by 
a hydrodynamical model with
first order phase transition. However, the description using the
transport equation allows to discuss alternative scenarios of
 the hadronization, e.g.  hadronization due to parton collisions in the
plasma \cite{NJL_had} with possible softening of the spectra of produced 
mesons. Another hadronization mechanism could be the fragmentation of
the fireball due to a possible instability of the system at finite
baryon density \cite{ins} or due to dynamical instabilities present for energy
densities corresponding to a mixed phase in the case of a first order
phase transition. The discussion of the hadronization mechanism and the
inclusion of the dynamics of mesons remains to be done.

\section*{Acknowledgments}
One of the authors (P.B.) wishes to thank the Alexander von Humboldt
Foundation for financial support. This work has been supported
in part by the German Ministry for Education and Research (BMBF)
under contract number 06 HD 742.

\end{document}